\documentclass[12pt]{iopart}

\usepackage{iopams}  
\usepackage{graphicx}
\usepackage{color}
\usepackage{soul}
\sethlcolor{yellow}

\begin{document}

\title{Nonequilibrium Work Relation in Macroscopic System}

\author{Yuki Sughiyama$^1$ and Masayuki Ohzeki$^{2,3}$}

\address{$^1$FIRST, Aihara Innovative Mathematical Modelling Project, Japan Science and Technology Agency, Institute of Industrial Science, The University of Tokyo, Komaba, Meguro-ku, Tokyo 153-8505, Japan\\
$^2$Department of Systems Science, Graduate School of Informatics, Kyoto University, Yoshida-Honmachi, Sakyo-ku, Kyoto, 606-8501, Japan\\
$^3$Dipartimento di Fisica, Universit\`a di Roma `La Sapienza', P.le Aldo Moro 2, Roma, 00185, Italy}
\ead{yuki$\underline{\ }$sughiyama@sat.t.u-tokyo.ac.jp}
\begin{abstract}
We reconsider a well-known relationship between the fluctuation theorem and the second law of thermodynamics by evaluating stochastic evolution of the density field (probability measure valued process).
In order to establish a bridge between microscopic and macroscopic behaviors, we must take the thermodynamic limit of a stochastic dynamical system following the standard procedure in statistical mechanics.
The thermodynamic path characterizing a dynamical behavior in the macroscopic scale can be formulated as an infimum of the action functional for the stochastic evolution of the density field.
In our formulation, the second law of thermodynamics can be derived only by symmetry of the action functional without recourse to the Jarzynski equality.
Our formulation leads to a nontrivial nonequilibrium work relation for metastable (quasi-stationary) states, which are peculiar in the macroscopic system.
We propose a prescription for computing the free energy for metastable states based on the resultant work relation.
\end{abstract}
\pacs{05.70.Ln, 05.10.Gg, 05.70.-a, 05.40.-a, 44.05.+e, 64.60.My}
\submitto{JSTAT}
\maketitle
\section{Introduction}
The Jarzynski equality \cite{01} $\left\langle e^{-\beta W}\right\rangle_{\mathrm{e}\mathrm{q}}=e^{-\beta\Delta\Phi_{\mathrm{e}\mathrm{q}}}$, where $\Delta\Phi_{\mathrm{e}\mathrm{q}}$ and $W$ are a free-energy difference and a performed work respectively, plays a role of a bridge between the symmetry of microscopic dynamics as given by the fluctuation theorem \cite{02,03,14,15,16,04,18}    and the fundamental limitation on time evolution of macroscopic quantities, which is described as the second law of thermodynamics.
The brackets $\left\langle\cdots\right\rangle_{\mathrm{eq}}$ denote the average over all the realizations during the dynamics starting from the initial equilibrium state.
The fluctuation theorem implies that the microscopic stochastic dynamics satisfies the following symmetry.
\begin{equation}
P\left[\lambda,\Gamma|x_{0}\right]e^{\beta Q\left[\lambda,\Gamma\right]}=P\left[\hat{\lambda},\hat{\Gamma}|x_{T}\right],\label{ppq}
\end{equation}
where $\lambda,\ \Gamma$, and $Q$ denote an external protocol, a microscopic time evolution of the system with the initial condition $x_{0}$, and a heat flow from the surrounding heat bath to the system, respectively.
In addition, $\hat{\cdot}$ stands for the time-reversal operation.
We usually term the Jarzynski equality as generalization of the second law of thermodynamics since the Jarzynski equality yields $\left\langle W\right\rangle_{\mathrm{eq}}\geq\Delta\Phi_{{\rm eq}}$ by use of Jensen's inequality.
Although the present study does not aim at giving any objection against the validity of the Jarzynski equality, we would like to ask the following fundamental question on the above ordinary understanding of the relationship between the Jarzynski equality and the second law of thermodynamics.
Does $\left\langle W\right\rangle_{\mathrm{eq}}$ express the performed work in thermodynamics?

The reasons for asking this question are as follows.
Let us go back to a starting point of statistical mechanics and its connection with thermodynamics.
To elucidate macroscopic properties from a microscopic description of the system, we consider an infinite-number limit of the sample mean, which is a coarse-grained picture of the system.
On the other hand, the performed work given in the Jarzynski equality $\left\langle W\right\rangle_{\mathrm{eq}}$ has not been baptized yet through this ordinary procedure.
Within a framework of the large deviation theory \cite{17,12},  which underlies statistical mechanics, the observed physical quantity in the thermodynamics is characterized by a global minimum of the rate function.
The performed work used in the Jarzynski equality is not yet confirmed to be expressed by the global minimum of the rate function.
In addition, the rate function can possess not only the global minimum but also several local minima, which represent existence of so-called metastable (quasi-stationary) states.
The system with metastable states does not always relax toward equilibrium state within a moderate time scale in a practical experiment.
From the above considerations, we cannot readily regard the performed work $\left\langle W\right\rangle_{\mathrm{eq}}$ in methods with the Jarzynski equality as that in thermodynamics.
This problem motivates us to compute the performed work in thermodynamics by directly evaluating the dynamics for the coarse-grained quantity, which can be described by stochastic evolution of the density field (probability measure valued process \cite{06,07}).

We analyze the stochastic evolution of the density field by employing the following strategy.
At first, we evaluate the entropy production of the system from the action functional \cite{12,07,10}, which is the rate function for the stochastic evolution of the density field and describes contributions around the most probable path of time evolution of the density fields.
As one of our main results in the present study, the second law of thermodynamics can be rederived from the symmetry of this action functional, which is generated from the fluctuation theorem.
This formalism is the baptized way toward thermodynamics by use of the rate function, which is different from the ordinary one through the Jarzynski equality as in the literatures \cite{03,15,16,04,18}.
These analyses provide deep understanding of the connection between the fluctuation theorem and the second law of thermodynamics and bring us another nontrivial result, which is relevant for macroscopic behavior in a system with metastable states.
The further analysis of the action functional leads the form of the rate function for the metastable states, which enables us to establish a nontrivial nonequilibrium work relation for metastable states.
This work relation can be expressed in the same form as the Jarzynski equality in the special condition related to relaxation processes toward the preselected metastable state denoted by an index $i$.
\begin{equation}
\left\langle e^{-N\beta w}\right\rangle_{i}=e^{-N\beta\left\{\phi_{{\rm eq}}\left(\beta,\lambda_{T}\right)-\phi_{i}\left(\beta,\lambda_{0}\right)\right\}},\label{JE_meta}
\end{equation}
where $N$ is the number of degrees of freedom in the system we deal with, $w$ is the performed work per degree of freedom through the external protocol characterized by $\lambda$.
The quantity $\phi_{{\rm eq}}$ is the free energy per degree of freedom for the last equilibrium state at $\lambda_{T}$, while $\phi_{i}$ is that for the initial metastable state identified by an index $i$ among several metastable states at $\lambda_{0}$.
By use of our nonequilibrium work relation (\ref{JE_meta}), we can compute the free energy for the metastable state due to the same form as the Jarzynski equality.
The existence of metastable states is considered to be related to peculiar relaxation dynamics as in the glassy systems.
Our result would provide a significant insight for the understanding of the special behavior observed in such complicated systems.

We organize the present paper as follows.
Before going to the detailed analysis, we explain our motivation in the present study in more detail by showing a gap of understanding on the fluctuation theorem from the description of the thermodynamic system.
In the following section, we demonstrate the detailed analysis on the action functional, which characterizes the rate function for the stochastic evolution of the density field.
The description of the second law of thermodynamics, by our presentation, is shown in Section 4.
In Section 5, we assess the Jarzynski equality for the system with metastable states.
In the last section, we conclude our study.

\section{Large deviation property and its importance}
In order to evaluate the most probable event in the stochastic system, we often evaluate the rate function.
In our study, we focus on the performed work in the thermodynamic system.
The Crooks fluctuation theorem \cite{03,14}   shows the symmetry of the path probability for the forward process and backward process with the performed work.
The Jarzynski equality can be obtained from one of the consequence of the Crooks fluctuation theorem.
One might think that all we have to do is to consider their limit of large-number degrees of freedom as $ N\to \infty$ (In Ref. \cite{Ohzeki2012}, the Crooks fluctuation theorem in context of the large deviation theory is presented).
It is however not always true.
The simple limitation is validated under a specific assumption that the large deviation property holds for the path probability associated with the performed work.
In other words, we assume existence of the rate function with a convex form for the performed work as
\begin{equation}
P(w)\sim\exp(-NI(w)),
\end{equation}
where $P(w)$ is the probability for the performed work, and $I(w)$ is its rate function.
The assumption is quite natural but often gives rise to our misunderstanding on the correct description of the macroscopic system.
For instance, the phase transition in equilibrium system can be explained in context of the rate function as its breakdown of convexity.
The form of the rate function should be discussed in detail to correctly understand thermodynamic behavior.

In the present study, we analyze the rate function without assumption on its form for a mean-field model.
In equilibrium statistical mechanics, the mean-field model such as the Curie-Weiss model plays a role to provide the essential understanding on the macroscopic behavior including the phase transition by describing various forms of the rate function of the order parameter.
We show the detailed analysis of the rate function in the following section.
After that, we point out possibility that a different nonequilibrium equality holds for the peculiar dynamics in the system with metastable states.

\section{Action functional for mean-field dynamics}
Let us consider that the system consists of $N$ degrees of freedom $\left\{x_{i}\right\}\left(i=1\cdots N\right)$ and is attached to a heat bath with an inverse temperature $\beta$.
The work is performed into the system through an external protocol $\lambda$.
Let us make two additional assumptions for the system in our study.
First, the dynamics is governed by the diffusion process \cite{05}. Second, the Hamiltonian is written by two-body interactions with all the components (i.e. mean field).
Under this assumption, the stochastic process satisfying the fluctuation theorem (\ref{ppq}) is described as
\begin{equation}
dx_{i}=\displaystyle \left[-\frac{\partial}{\partial x_{i}}H_{N}\left(\left\{x_{i}\right\},\lambda_{t}\right)\right]dt+\sqrt{\frac{2}{\beta}}d\xi_{i},\label{eq3}
\end{equation}
where $\xi_{i}$ denotes the Wiener process, and $H_{N}\left(\left\{x_{i}\right\},\lambda_{t}\right)$ is the Hamiltonian of the system. 
The Hamiltonian is written by 
\begin{equation}
H_{N}\displaystyle \left(\left\{x_{i}\right\},\lambda\right)=-\sum_{i}F_{\lambda}\left(x_{i}\right)-\frac{1}{2N}\sum_{ij}G\left(x_{i},x_{j}\right),\label{mfhm}
\end{equation}
where $F_{\lambda}$ stands for the potential trap, $G$ is the two-body interaction term and symmetric $G\left(x,y\right)=G\left(y,x\right)$.
In this dynamical system, we analyze stochastic evolution of the density field (empirical measure) $\mu\left(x,t\right)=\left(1/N\right)\Sigma_{i}\delta\left(x_{i}-x\right)$ \cite{06,07,08,09}.
By employing equation (\ref{eq3}) and the Ito formula, we evaluate the derivative of $\mu$ as
\begin{eqnarray}
\displaystyle \nonumber d\mu\left(x,t\right)&=&\displaystyle \sum_{j}\frac{\partial}{\partial x_{j}}\left\{\frac{1}{N}\sum_{i}\delta\left(x_{i}-x\right)\right\}dx_{j}\\
\displaystyle \nonumber&&+\frac{1}{2}\sum_{k,l}\frac{\partial^{2}}{\partial x_{k}\partial x_{l}}\left\{\frac{1}{N}\sum_{i}\delta\left(x_{i}-x\right)\right\}dx_{k}dx_{l}\\
\displaystyle \nonumber&=&\displaystyle \frac{1}{N}\sum_{i}\frac{\partial\delta\left(x_{i}-x\right)}{\partial x_{i}}dx_{i}+\frac{1}{N\beta}\sum_{i}\frac{\partial^{2}\delta\left(x_{i}-x\right)}{\partial x_{i}^{2}}dt\\
\displaystyle \nonumber&=&\displaystyle \frac{1}{N}\sum_{i}\frac{\partial\delta\left(x_{i}-x\right)}{\partial x_{i}}\left\{f_{\lambda}\left(x_{i}\right)+\frac{1}{N}\sum_{j}g\left(x_{i},x_{j}\right)\right\}dt\\
&&+\displaystyle \frac{1}{N\beta}\sum_{i}\frac{\partial^{2}\delta\left(x_{i}-x\right)}{\partial x_{i}^{2}}dt+\frac{1}{N}\sqrt{\frac{2}{\beta}}\sum_{i}\frac{\partial\delta\left(x_{i}-x\right)}{\partial x_{i}}d\xi_{i},\label{eq4}
\end{eqnarray}
where we use abbreviations $f_{\lambda_{t}}\left(x\right)=\partial F_{\lambda_{t}}\left(x\right)/\partial x,\ g\left(x,y\right)=\partial G\left(x,y\right)/\partial x$ and the property of the Wiener process $d\xi_{k}d\xi_{l}=\delta_{k,l}dt$, where $\delta_{k,l}$ denotes Kronecker delta.
We substitute properties of the density field,
\begin{eqnarray}
\displaystyle \frac{\partial^{2}\mu\left(x,t\right)}{\partial x^{2}}=\frac{1}{N}\sum_{i}\frac{\partial^{2}\delta\left(x_{i}-x\right)}{\partial x_{i}^{2}},\\
\displaystyle \frac{\partial}{\partial x}\left\{A\left(x\right)\mu\left(x,t\right)\right\}=-\frac{1}{N}\sum_{i}A\left(x_{i}\right)\frac{\partial\delta\left(x_{i}-x\right)}{\partial x_{i}},\\
\displaystyle \int A\left(x\right)\mu\left(x,t\right)dx=\frac{1}{N}\sum_{i}A\left(x_{i}\right),
\end{eqnarray}
where $A\left(x\right)$ is an arbitrary function, into equation (\ref{eq4}).
Then, we obtain the stochastic evolution of the density field as
\begin{equation}
d\displaystyle \mu\left(x,t\right)=D_{\mu_{t},\lambda_{t}}\left(x\right)dt+\frac{1}{N}\sqrt{\frac{2}{\beta}}\sum_{i}\frac{\partial\delta\left(x_{i}-x\right)}{\partial x_{i}}d\xi_{i},\label{dmu}
\end{equation}
where
\begin{eqnarray}
\displaystyle \nonumber D_{\mu_{t},\lambda_{t}}\left(x\right)&=&-\displaystyle \frac{\partial}{\partial x}\left\{f_{\lambda_{t}}\left(x\right)\mu\left(x,t\right)\right\}\\
&&-\displaystyle \frac{\partial}{\partial x}\left\{\mu\left(x,t\right)\int g\left(x,y\right)\mu\left(y,t\right)dy\right\}+\frac{1}{\beta}\frac{\partial^{2}\mu\left(x,t\right)}{\partial x^{2}}.\label{xxx68}
\end{eqnarray}
By employing equation (\ref{dmu}), we derive the functional Fokker-Planck equation as, following the ordinary procedure \cite{05}     [see also Appendix A],
\begin{eqnarray}
\displaystyle \nonumber\frac{\partial}{\partial t}\mathrm{Prob}_{t}\left[\mu\right]&=&\displaystyle \int dx\frac{\delta}{\delta\mu\left(x\right)}\left\{-D_{\mu,\lambda_{t}}\left(x\right)\mathrm{Prob}_{t}\left[\mu\right]\right\}\\
&&+\displaystyle \frac{1}{2N}\int\int dxdy\frac{\delta^{2}}{\delta\mu\left(x\right)\delta\mu\left(y\right)}\left\{R_{\mu}\left(x,y\right)\mathrm{Prob}_{t}\left[\mu\right]\right\},\label{fokker-planck}
\end{eqnarray}
where $\mathrm{Prob}_{t}[\mu]$ is the probability for the density field and $R_{\mu}\left(x,y\right)$ denotes a diffusion matrix of the functional Fokker-Planck equation (\ref{fokker-planck}) as
\begin{equation}
R_{\mu}\displaystyle \left(x,y\right)=\frac{2}{\beta}\int dz\frac{\partial\delta\left(x-z\right)}{\partial z}\frac{\partial\delta\left(y-z\right)}{\partial z}\mu\left(z\right).\label{eq7}
\end{equation}

Let us consider the thermodynamic limit ($ N\rightarrow\infty$) on equation (\ref{dmu}) to elucidate the macroscopic behavior of the system.
We reach a nonlinear diffusion equation for the density field $\mu(x,t)$, which represents thermodynamic time evolution of the system,
\begin{equation}
\displaystyle \frac{\partial\mu\left(x,t\right)}{\partial t}=D_{\mu_{t},\lambda_{t}}\left(x\right).\label{diffusion eq}
\end{equation}

Next let us consider to derive the second law of thermodynamics through the fluctuation theorem.
In order to examine the fluctuation of the density field from the solution of the nonlinear diffusion equation (\ref{diffusion eq}) in a sufficient large $N$, we analyze the conditional path probability for the stochastic dynamics governed by the functional Fokker-Planck equation (\ref{fokker-planck}).
In the present case, we can express the conditional path probability during the time interval $\left[0,T\right]$ by the exponential form as $\mathrm{Prob}\left[\lambda,\mu|\mu_{0}\right]=e^{-NJ_{\left[0,T\right]}\left[\lambda,\mu\right]}$.
In other words, we find the rate function without assumption on its form.
Here, $J_{\left[0,T\right]}\left[\lambda,\mu\right]$ is called the action functional given as,
\begin{equation}
J_{\left[0,T\right]}\displaystyle \left[\lambda,\mu\right]=\int_{0}^{T}dtL_{\lambda_{t}}\left[\dot{\mu}_{\mathrm{t}},\mu_{t}\right]\label{J},
\end{equation}
where $L_{\lambda_{t}}\left[\dot{\mu}_{\mathrm{t}},\mu_{t}\right]$ is obtained by using equation (\ref{fokker-planck}) as \cite{12,10,11}
\begin{eqnarray}
\displaystyle \nonumber L_{\lambda_{t}}\left[\dot{\mu}_{t},\mu_{t}\right]&=&\displaystyle \frac{1}{2}\int\int dxdyR_{\mu_{t}}^{-1}\left(x,y\right)\left\{\dot{\mu}\left(x,t\right)-D_{\mu_{t},\lambda_{t}}\left(x\right)\right\}\\
&&\times\left\{\dot{\mu}\left(y,t\right)-D_{\mu_{t},\lambda_{t}}\left(y\right)\right\}.
\end{eqnarray}
Here, $R_{\mu_{t}}^{-1}\left(x,y\right)$ denotes the inverse of the diffusion matrix, which satisfies the identity
\begin{equation}
\displaystyle \int d\alpha R_{\mu_{t}}\left(x,\alpha\right)R_{\mu_{t}}^{-1}\left(\alpha,y\right)=\delta\left(x-y\right).\label{xxx67}
\end{equation}
For subsequent calculations, we transform equation (\ref{xxx67}) to a useful form. 
By substituting equation (\ref{eq7}) into equation (\ref{xxx67}), we obtain 
\begin{eqnarray}
\displaystyle \nonumber\delta\left(x-y\right)&=&\displaystyle \frac{2}{\beta}\int d\alpha\int dz\frac{\partial\delta\left(z-x\right)}{\partial z}\frac{\partial\delta\left(z-\alpha\right)}{\partial z}\mu\left(z,t\right)R_{\mu_{t}}^{-1}\left(\alpha,y\right)\\
\displaystyle \nonumber&=&-\displaystyle \frac{2}{\beta}\int d\alpha\int dz\delta\left(z-\alpha\right)R_{\mu_{t}}^{-1}\left(\alpha,y\right)\frac{\partial}{\partial z}\left\{\mu\left(z,t\right)\frac{\partial\delta\left(z-x\right)}{\partial z}\right\}\\
\displaystyle \nonumber&=&-\displaystyle \frac{2}{\beta}\int dzR_{\mu_{t}}^{-1}\left(z,y\right)\frac{\partial}{\partial z}\left\{\mu\left(z,t\right)\frac{\partial\delta\left(z-x\right)}{\partial z}\right\}\\
&=&-\displaystyle \frac{2}{\beta}\int dz\delta\left(z-x\right)\frac{\partial}{\partial z}\left\{\mu\left(z,t\right)\frac{\partial R_{\mu_{t}}^{-1}\left(z,y\right)}{\partial z}\right\}.
\end{eqnarray}
Thus, we find that the inverse of the diffusion matrix $R_{\mu_{t}}^{-1}$ satisfies the following equation,
\begin{equation}
\displaystyle \frac{\partial}{\partial x}\left\{\mu\left(x,t\right)\frac{\partial R_{\mu_{t}}^{-1}\left(x,y\right)}{\partial x}\right\}=-\frac{\beta}{2}\delta\left(x-y\right).\label{r}
\end{equation}
Owing to this property, we can reveal the symmetry of the action functional corresponding to the fluctuation theorem given by the ratio of the time forward path probability written by $L_{\lambda_{t}}\left[\dot{\mu}_{t},\mu_{t}\right]$ to the time backward one by $L_{\hat{\lambda}_{\hat{t}}}\left[\dot{\hat{\mu}}_{\hat{t}},\hat{\mu}_{\hat{t}}\right]$. Here, $\hat{\cdot}$ stands for the time-reversal operation: $\hat{\mu}_{\hat{t}}=\mu_{t}, \hat{\lambda}_{\hat{t}}=\lambda_{t}, \hat{t}=T-t$. In particular, the time derivation of $\hat{\mu}_{\hat{t}}$ is $\dot{\hat{\mu}}_{\hat{t}}=-\dot{\mu}_{t}$, where we use $\partial/\partial\hat{t}=-\partial/\partial t$.
This ratio is indeed evaluated as,
\begin{eqnarray}
\displaystyle \nonumber L_{\lambda_{t}}\left[\dot{\mu}_{t},\mu_{t}\right]&=&\displaystyle \frac{1}{2}\int\int dxdyR_{\hat{\mu}_{\hat{\mathrm{t}}}}^{-1}\left(x,y\right)\left\{-\dot{\hat{\mu}}\left(x,\hat{t}\right)-D_{\hat{\mu}_{\hat{t}},\hat{\lambda}_{\hat{\mathrm{t}}}}\left(x\right)\right\}\\
\nonumber&&\times\left\{-\dot{\hat{\mu}}\left(y,\hat{t}\right)-D_{\hat{\mu}_{\hat{t}},\hat{\lambda}_{\hat{\mathrm{t}}}}\left(y\right)\right\}\\
\displaystyle \nonumber&=&\displaystyle \frac{1}{2}\int\int dxdyR_{\hat{\mu}_{\hat{t}}}^{-1}\left(x,y\right)\left\{\dot{\hat{\mu}}\left(x,\hat{t}\right)-D_{\hat{\mu}_{\hat{t}},\hat{\lambda}_{\hat{\mathrm{t}}}}\left(x\right)\right\}\\
\nonumber&&\times\left\{\dot{\hat{\mu}}\left(y,\hat{t}\right)-D_{\hat{\mu}_{\hat{t}},\hat{\lambda}_{\hat{\mathrm{t}}}}\left(y\right)\right\}\\
\displaystyle \nonumber&&\ \ \ \ -\int\int dxdyR_{\hat{\mu}_{\hat{t}}}^{-1}\left(x,y\right)\dot{\hat{\mu}}\left(x,\hat{t}\right)\left\{\dot{\hat{\mu}}\left(y,\hat{t}\right)-D_{\hat{\mu}_{\hat{t}},\hat{\lambda}_{\hat{\mathrm{t}}}}\left(y\right)\right\}\\
\displaystyle \nonumber&&\ \ \ \ \ \ \ \ -\int\int dxdyR_{\hat{\mu}_{\hat{t}}}^{-1}\left(x,y\right)\left\{\dot{\hat{\mu}}\left(x,\hat{t}\right)-D_{\hat{\mu}_{\hat{t}},\hat{\lambda}_{\hat{\mathrm{t}}}}\left(x\right)\right\}\dot{\hat{\mu}}\left(y,\hat{t}\right)\\
\displaystyle \nonumber&&\ \ \ \ \ \ \ \ \ \ \ \ +2\int\int dxdyR_{\hat{\mu}_{\hat{t}}}^{-1}\left(x,y\right)\dot{\hat{\mu}}\left(x,\hat{t}\right)\dot{\hat{\mu}}\left(y,\hat{t}\right)\\
&=&L_{\hat{\lambda}_{\hat{t}}}\displaystyle \left[\dot{\hat{\mu}}_{\hat{t}},\hat{\mu}_{\hat{t}}\right]+2\int\int dxdyR_{\hat{\mu}_{\hat{t}}}^{-1}\left(x,y\right)D_{\hat{\mu}_{\hat{t}},\hat{\lambda}_{\hat{\mathrm{t}}}}\left(x\right)\dot{\hat{\mu}}\left(y,\hat{t}\right),
\end{eqnarray}
where we use the symmetric property $R_{\hat{\mu}_{\hat{t}}}^{-1}\left(x,y\right)=R_{\hat{\mu}_{\hat{t}}}^{-1}\left(y,x\right)$.
Thus, we obtain 
\begin{equation}
L_{\lambda_{t}}\displaystyle \left[\dot{\mu}_{t},\mu_{t}\right]=L_{\hat{\lambda}_{\hat{t}}}\left[\dot{\hat{\mu}}_{\hat{t}},\hat{\mu}_{\hat{t}}\right]-2\int\int dxdyR_{\mu_{t}}^{-1}\left(x,y\right)D_{\mu_{t},\lambda_{t}}\left(x\right)\dot{\mu}\left(y,t\right).\label{rtime}
\end{equation}
The explicit form of $D_{\mu_{t},\lambda_{t}}$ [see equation (\ref{xxx68})] enables us to calculate the second term of the right hand side in equation (\ref{rtime}) by use of integration by parts as 
\begin{eqnarray}
\displaystyle \nonumber\int\int dxdyR_{\mu_{t}}^{-1}\left(x,y\right)D_{\mu_{t},\lambda_{t}}\left(x\right)\dot{\mu}\left(y,t\right)\\
\displaystyle \nonumber=-\int\int dxdy\frac{\partial R_{\mu_{t}}^{-1}\left(x,y\right)}{\partial x}\\
\displaystyle \nonumber\ \ \ \ \times\left\{-f_{\lambda_{t}}\left(x\right)\mu\left(x,t\right)-\mu\left(x,t\right)\int g\left(x,z\right)\mu\left(z,t\right)dz+\frac{1}{\beta}\frac{\partial\mu\left(x,t\right)}{\partial x}\right\}\dot{\mu}\left(y,t\right)\\
\displaystyle \nonumber=-\int\int dxdy\frac{\partial R_{\mu_{t}}^{-1}\left(x,y\right)}{\partial x}\mu\left(x,t\right)\\
\displaystyle \nonumber\ \ \ \ \times\left\{-\frac{\partial F_{\lambda_{t}\left(x\right)}}{\partial x}-\frac{\partial}{\partial x}\int G\left(x,z\right)\mu\left(z,t\right)dz+\frac{1}{\beta}\frac{\partial\log\mu\left(x,t\right)}{\partial x}\right\}\dot{\mu}\left(y,t\right)\\
\displaystyle \nonumber=\int\int dxdy\frac{\partial}{\partial x}\left\{\mu\left(x,t\right)\frac{\partial R_{\mu_{t}}^{-1}\left(x,y\right)}{\partial x}\right\}\\
\displaystyle \ \ \ \ \times\left\{-F_{\lambda_{t}}\left(x\right)-\int G\left(x,z\right)\mu\left(z,t\right)dz+\frac{1}{\beta}\log\mu\left(x,t\right)\right\}\dot{\mu}\left(y,t\right).
\end{eqnarray}
In addition, by employing the property of the inverse diffusion matrix (\ref{r}), we obtain 
\begin{eqnarray}
\displaystyle \nonumber\int\int dxdyR_{\mu_{t}}^{-1}\left(x,y\right)D_{\mu_{t},\lambda_{t}}\left(x\right)\dot{\mu}\left(y,t\right)\\
\displaystyle \nonumber=-\frac{\beta}{2}\int\int dxdy\delta\left(x-y\right)\\
\displaystyle \nonumber\ \ \ \ \times\left\{-F_{\lambda_{t}}\left(x\right)-\int G\left(x,z\right)\mu\left(z,t\right)dz+\frac{1}{\beta}\log\mu\left(x,t\right)\right\}\dot{\mu}\left(y,t\right)\\
\displaystyle \nonumber=-\frac{1}{2}\int dx\left\{-\beta F_{\lambda_{t}}\left(x\right)-\beta\int G\left(x,z\right)\mu\left(z,t\right)dz+\log\mu\left(x,t\right)\right\}\dot{\mu}\left(x,t\right).\\\label{xxxx60}
\end{eqnarray}
On the other hand, we evaluate the variation of the rate function for the canonical distribution $\exp(-\beta H_{N})/Z_{N}$ with the external parameter $\lambda_{t}$, which is $I_{{\rm eq},\lambda_{t}}\left[\mu_{t}\right]=\beta\epsilon_{\lambda_{t}}\left[\mu_{t}\right]-s\left[\mu_{t}\right]-\beta\phi_{{\rm eq}}\left(\beta,\lambda_{t}\right)$, where $\epsilon_{\lambda},\ s$ and $\phi_{{\rm eq}}$ represent the energy per degree of freedom, the same form as the Shannon entropy and the free energy per degree of freedom, respectively \cite{12}     [see also Appendix B]. 
Then, we obtain 
\begin{equation}
\displaystyle \frac{\delta I_{{\rm eq},\lambda_{t}}\left[\mu_{t}\right]}{\delta\mu\left(x,t\right)}=-\beta F_{\lambda_{t}}\left(x\right)-\beta\int G\left(x,z\right)\mu\left(z,t\right)dz+\log\mu\left(x,t\right)+1.\label{xxxx65}
\end{equation}
where we use the explicit form of $\epsilon_{\lambda}$ and $s$,
\begin{eqnarray}
\displaystyle \epsilon_{\lambda}\left[\mu\right]=-\int F_{\lambda}\left(x\right)\mu\left(x\right)dx-\frac{1}{2}\int\mu\left(x\right)G\left(x,y\right)\mu\left(y\right)dxdy,\\
s\displaystyle \left[\mu\right]=-\int\mu\left(x\right)\log\mu\left(x\right)dx.
\end{eqnarray}
By multiplying both sides in equation (\ref{xxxx65}) by $\dot{\mu}\left(x,t\right)$ and integrating them, we find
\begin{eqnarray}
\displaystyle \nonumber\int dx\frac{\delta I_{{\rm eq},\lambda_{t}}\left[\mu_{t}\right]}{\delta\mu\left(x,t\right)}\dot{\mu}\left(x,t\right)\\
=\displaystyle \int dx\left\{-\beta F_{\lambda_{t}}\left(x\right)-\beta\int G\left(x,z\right)\mu\left(z,t\right)dz+\log\mu\left(x,t\right)\right\}\dot{\mu}\left(x,t\right),\label{xxxx66}
\end{eqnarray}
where we use the normalization condition $\displaystyle \int\dot{\mu}\left(x,t\right)dx=0$.
Comparison of equation (\ref{xxxx60}) with equation (\ref{xxxx66}) enables us to evaluate equation (\ref{rtime}) as
\begin{equation}
L_{\lambda_{t}}\displaystyle \left[\dot{\mu}_{t},\mu_{t}\right]=L_{\hat{\lambda}_{\hat{t}}}\left[\dot{\hat{\mu}}_{\hat{t}},\hat{\mu}_{\hat{t}}\right]+\int dx\frac{\delta I_{{\rm eq},\lambda_{t}}\left[\mu_{t}\right]}{\delta\mu\left(x,t\right)}\dot{\mu}\left(x,t\right).\label{xxxx67}
\end{equation}
By substituting equation (\ref{xxxx67}) into the action functional (\ref{J}), we obtain that the symmetry of action functional $J_{\left[0,T\right]}\left[\lambda,\mu\right]$ can be written as
\begin{equation}
J_{\left[0,T\right]}\displaystyle \left[\lambda,\mu\right]=J_{\left[0,T\right]}\left[\hat{\lambda},\hat{\mu}\right]+\int_{0}^{T}dt\left(\frac{dI_{{\rm eq},\lambda_{t}}\left[\mu_{t}\right]}{dt}-\frac{\partial I_{{\rm eq},\lambda_{t}}\left[\mu_{t}\right]}{\partial\lambda_{t}}\dot{\lambda}_{t}\right).\label{xxxx68}
\end{equation}
Here, we note that the second term of the right hand side in equation (\ref{xxxx68}) is represented as
\begin{eqnarray}
\displaystyle \nonumber\int_{0}^{T}dt\left(\frac{dI_{{\rm eq},\lambda_{t}}\left[\mu_{t}\right]}{dt}-\frac{\partial I_{{\rm eq},\lambda_{t}}\left[\mu_{t}\right]}{\partial\lambda_{t}}\dot{\lambda}_{t}\right)\\
\displaystyle \nonumber=I_{{\rm eq},\lambda_{T}}\left[\mu_{T}\right]-I_{{\rm eq},\lambda_{0}}\left[\mu_{0}\right]-\beta\int_{0}^{T}dt\frac{\partial\epsilon_{\lambda_{t}}\left[\mu_{t}\right]}{\partial\lambda_{t}}\dot{\lambda}_{t}\\
\ \ \ \ +\beta\left\{\phi_{{\rm eq}}\left(\beta,\lambda_{T}\right)-\phi_{{\rm eq}}\left(\beta,\lambda_{0}\right)\right\},
\end{eqnarray}
and the work per degree of freedom performed by the external protocol $\lambda$ during the time interval $\left[0,T\right]$ is written by 
\begin{equation}
w_{\left[0,T\right]}\displaystyle \left[\lambda,\mu\right]=\int_{0}^{T}dt\frac{\partial\epsilon_{\lambda_{t}}\left[\mu_{t}\right]}{\partial\lambda_{t}}\dot{\lambda}_{t}.
\end{equation}
Taking these facts into account, we find that the symmetry of action functional (\ref{xxxx68}) is represented as
\begin{eqnarray}
\nonumber&& I_{{\rm eq},\lambda_{0}}\left[\mu_{0}\right]+J_{\left[0,T\right]}\left[\lambda,\mu\right]+\beta w_{\left[0,T\right]}\left[\lambda,\mu\right]\\
&&=I_{{\rm eq},\hat{\lambda}_{0}}\left[\hat{\mu}_{0}\right]+J_{\left[0,T\right]}\left[\hat{\lambda},\hat{\mu}\right]+\beta\left\{\phi_{{\rm eq}}\left(\beta,\lambda_{T}\right)-\phi_{{\rm eq}}\left(\beta,\lambda_{0}\right)\right\}.\label{fluctuation theorem}
\end{eqnarray}
Equation (\ref{fluctuation theorem}) stands for the fluctuation theorem described by the action functional and allows us to formulate the second law of thermodynamics in terms of the work as usually observed in thermodynamics.

\section{The second law of thermodynamics}
Now we are in a position to end the discussion on the second law of thermodynamics starting from the coarse-grained picture of the system. 
In terms of the action functional, the thermodynamic path $\mu^{*}$, which denotes a solution of the nonlinear diffusion equation (\ref{diffusion eq}), must satisfy 
$J_{\left[0,T\right]}\displaystyle \left[\lambda,\mu^{*}\right]=\inf_{\mu}J_{\left[0,T\right]}\left[\lambda,\mu\right]$, i.e. $J_{\left[0,T\right]}\left[\lambda,\mu^{*}\right]=0$.
The second law of thermodynamics is written in the inequality form, which connects the performed work and the free-energy difference between two different equilibrium states. 
We thus choose the initial condition of equation (\ref{diffusion eq}) as the equilibrium state $\mu_{{\rm eq},\lambda_{0}}$, 
which holds $I_{{\rm eq},\lambda_{0}}\left[\mu_{{\rm eq},\lambda_{0}}\right]=0$. 
The thermodynamic path launched from the equilibrium state, $\mu^{*}\left[\lambda;\mu_{{\rm eq},\lambda_{0}}\right]$, which means the most probable path, satisfies 
\begin{equation}
I_{{\rm eq},\lambda_{0}}\left[\mu_{{\rm eq},\lambda_{0}}\right]+J_{\left[0,T\right]}\left[\lambda,\mu^{*}\left[\lambda;\mu_{{\rm eq},\lambda_{0}}\right]\right]=0.\label{ji}
\end{equation}
Substituting equation (\ref{ji}) into the symmetry (\ref{fluctuation theorem}), we obtain 
\begin{eqnarray}
\nonumber\beta w_{\left[0,T\right]}\left[\lambda,\mu^{*}\left[\lambda;\mu_{{\rm eq},\lambda_{0}}\right]\right]\\
\nonumber=I_{{\rm eq},\hat{\lambda}_{0}}\left[\hat{\mu}_{0}^{*}\left[\lambda;\mu_{{\rm eq},\lambda_{0}}\right]\right]+J_{\left[0,T\right]}\left[\hat{\lambda},\hat{\mu}^{*}\left[\lambda;\mu_{{\rm eq},\lambda_{0}}\right]\right]\\
\ \ \ \ +\beta\left\{\phi_{{\rm eq}}\left(\beta,\lambda_{T}\right)-\phi_{{\rm eq}}\left(\beta,\lambda_{0}\right)\right\},
\end{eqnarray}
where $\hat{\mu}^{*}\left[\lambda;\mu_{{\rm eq},\lambda_{0}}\right]$ denotes the time-reversal most probable path,   $\hat{\mu}_{0}^{*}\left[\lambda;\mu_{{\rm eq},\lambda_{0}}\right]$ is its initial condition and $w\left[\lambda,\mu^{*}\left[\lambda;\mu_{{\rm eq},\lambda_{0}}\right]\right]$ stands for the work performed on the system which is termed in thermodynamics.
Since the action functional $J_{\left[0,T\right]}\left[\lambda,\mu\right]$ and the rate function $I_{{\rm eq},\lambda}\left[\mu\right]$ always take a non-negative value, $I_{{\rm eq},\hat{\lambda}_{0}}\left[\hat{\mu}_{0}^{*}\left[\lambda;\mu_{{\rm eq},\lambda_{0}}\right]\right]+J_{\left[0,T\right]}\left[\hat{\lambda},\hat{\mu}^{*}\left[\lambda;\mu_{{\rm eq},\lambda_{0}}\right]\right]\geq 0$, we find the second law of thermodynamics $w_{\left[0,T\right]}\left[\lambda,\mu^{*}\left[\lambda;\mu_{{\rm eq},\lambda_{0}}\right]\right]\geq\phi_{{\rm eq}}\left(\beta,\lambda_{T}\right)-\phi_{{\rm eq}}\left(\beta,\lambda_{0}\right)$. 
The entropy production $\sigma\left[\lambda\right]=\beta\left\{w_{\left[0,T\right]}\left[\lambda,\mu^{*}\left[\lambda;\mu_{{\rm eq},\lambda_{0}}\right]\right]-\Delta\phi_{\rm eq}\right\}$ can also be written in terms of the action functional and the rate function as
\begin{equation}
\sigma\left[\lambda\right]=I_{{\rm eq},\hat{\lambda}_{0}}\left[\hat{\mu}_{0}^{*}\left[\lambda;\mu_{{\rm eq},\lambda_{0}}\right]\right]+J_{\left[0,T\right]}\left[\hat{\lambda},\hat{\mu}^{*}\left[\lambda;\mu_{{\rm eq},\lambda_{0}}\right]\right].\label{fdr}
\end{equation}

Before closing this section, we show the derivation of the ordinary Jarzynski equality from the symmetry (\ref{fluctuation theorem}). 
Suppose that the system is set in an initial equilibrium state. 
Then, the rate function for the time evolution of the density field is given by $I_{{\rm eq},\lambda_{0}}\left[\mu_{0}\right]+J_{\left[0,T\right]}\left[\lambda,\mu\right]$.
We can evaluate the expectation of the exponentiated work in a sufficient large $N$ (i.e. $ N\rightarrow\infty$) as, by employing Varadhan's theorem \cite{13} (also see equation (60) in section 3.5.4 of the reference [10]),
\begin{equation}
-\displaystyle \frac{1}{N}\log\left\langle e^{-N\beta w}\right\rangle_{\rm eq}=\inf_{\mu}\left(I_{{\rm eq},\lambda_{0}}\left[\mu_{0}\right]+J_{\left[0,T\right]}\left[\lambda,\mu\right]+\beta w_{\left[0,T\right]}\left[\lambda,\mu\right]\right).
\end{equation}
Here, using the symmetry (\ref{fluctuation theorem}), we obtain 
\begin{eqnarray}
\nonumber
-\displaystyle \frac{1}{N}\log\left\langle e^{-N\beta w}\right\rangle_{{\rm eq}}&=&\displaystyle \beta\Delta\phi_{{\rm eq}}+\inf_{\hat{\mu}}\left(I_{{\rm eq},\hat{\lambda}_{0}}\left[\hat{\mu}_{0}\right]+J_{\left[0,T\right]}\left[\hat{\lambda},\hat{\mu}\right]\right)\\
&=&\beta\Delta\phi_{{\rm eq}},
\end{eqnarray}
where $\Delta\phi_{{\rm eq}}=\phi_{{\rm eq}}\left(\beta,\lambda_{T}\right)-\phi_{{\rm eq}}\left(\beta,\lambda_{0}\right)$ and we use the property of the rate function, $\displaystyle \inf_{\hat{\mu}}\left(I_{{\rm eq},\hat{\lambda}_{0}}\left[\hat{\mu}_{0}\right]+J_{\left[0,T\right]}\left[\hat{\lambda},\hat{\mu}\right]\right)=0$.
Accordingly, we find the ordinary Jarzynski equality $\left\langle e^{-N\beta w}\right\rangle_{{\rm eq}}=e^{-N\beta\Delta\phi_{{\rm eq}}}$.
\section{Nonequilibrium work relation for metastable states}
\begin{figure}[h]
\begin{center}
\includegraphics*[height=5cm]{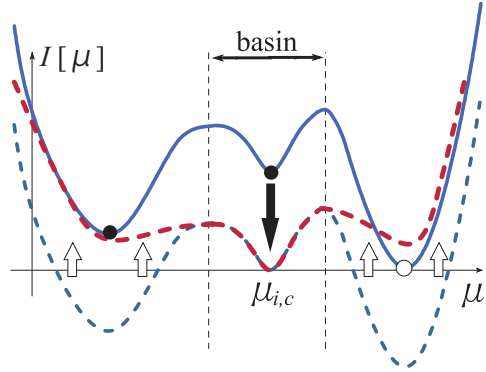}
\caption{(Color online) The blue curve represents the Lyapunov function of the nonlinear diffusion equation (the rate function for equilibrium state), $I_{{\rm eq},c}$. 
The red dashed curve denotes the rate function for the $i$th metastable state, $V_{i,c}$.
The black circles represent the metastable states, while the white circle denote the equilibrium state.
The black arrow represents the downward parallel translation by $I_{{\rm eq},c}\left[\mu_{i,c}\right]$ of the rate function $I_{{\rm eq},c}$ as the blue dashed curve.
The white arrows describe the modification by $A_{i,c}$ of the blue dashed curve into the red dashed curve $V_{i,c}$.}
\label{fig1}
\end{center}
\end{figure}

We give remarks on several properties of metastable states.
First, we show how metastable states emerge in relaxation process. 
Suppose that the system is not perturbed during relaxation process.
We choose the external parameter $\lambda$ as constant $c$.
The thermodynamic time evolution is given by the nonlinear diffusion equation (\ref{diffusion eq}).
Thus, metastable states and equilibrium state are given as fixed points of this equation. 
In order to obtain the fixed points of equation (\ref{diffusion eq}), we examine the Lyapunov function \cite{05}. 
We find that the rate function $I_{{\rm eq},c}\left[\mu\right]$ plays a roll of the Lyapunov function as follows.
First, the rate function $I_{{\rm eq},c}\left[\mu\right]$ satisfies $I_{{\rm eq},c}\left[\mu\right]\geq 0$ for all $\mu$. 
Second, the derivative of $I_{{\rm eq},c}\left[\mu\right]$ with $t$ is represented as
\begin{eqnarray}
\displaystyle \nonumber\frac{dI_{{\rm eq},c}\left[\mu\right]}{dt}&=&\displaystyle \int dx\frac{\delta I_{{\rm eq},c}\left[\mu\right]}{\delta\mu\left(x,t\right)}\dot{\mu}\left(x,t\right)\\
\displaystyle \nonumber&=&-\displaystyle \beta\int dx\mu\left(x,t\right)\\
&&\displaystyle \times\left\{f_{c}\left(x\right)+\int g\left(x,y\right)\mu\left(y,t\right)dy-\frac{1}{\beta}\frac{\partial\log\mu\left(x,t\right)}{\partial x}\right\}^{2}\leq 0,
\end{eqnarray}
where we use equation (27) with equations (11) and (14).
Thus, we can regard the rate function $I_{{\rm eq},c}\left[\mu\right]$ as the Lyapunov function of the nonlinear diffusion equation (\ref{diffusion eq}).
From this fact, the metastable states and equilibrium state are given as local minima and a global minimum of $I_{{\rm eq},c}\left[\mu\right]$, respectively [see Figure \ref{fig1}].
Second, let us show the rate function for metastable states. 
Suppose that the system has $n$ metastable states.
We then focus on the $i$th metastable state denoted by $\mu_{i,c}$. 
The rate function for this metastable state can be represented as \cite{12,07,10}     
\begin{equation}
V_{i,c}\displaystyle \left[\nu\right]=\inf_{\mu:\mu_{0}=\mu_{i,c},\mu_{\infty}=\nu}J_{\left[0,\infty\right]}^{c}\left[\mu\right],\label{quasi-potential}
\end{equation}
where $\displaystyle \inf_{\mu:\mu_{0}=\mu_{i,c},\mu_{\infty}=\nu}$ represents an infimum among all the paths $\mu$ launched from $\mu_{i,c}$ to $\nu$ during the time interval $\left[0,\infty\right]$. 
Here, $\nu$ represents the density field after the relaxation.
To evaluate the rate function (\ref{quasi-potential}), we employ the constant protocol version of the symmetry (\ref{fluctuation theorem}), 
\begin{equation}
I_{{\rm eq},c}\left[\mu_{0}\right]+J_{\left[0,T\right]}^{c}\left[\mu\right]=I_{{\rm eq},c}\left[\hat{\mu}_{0}\right]+J_{\left[0,T\right]}^{c}\left[\hat{\mu}\right].
\end{equation}
Then, we obtain the rate function for the metastable state as \cite{07,11} [see also Figure \ref{fig1}], 
\begin{eqnarray}
\displaystyle \nonumber V_{i,c}\left[\nu\right]&=&\displaystyle \inf_{\mu:\mu_{0}=\mu_{i,c},\mu_{\infty}=\nu}J_{\left[0,\infty\right]}^{c}\left[\mu\right]\\
\displaystyle \nonumber&=&\displaystyle \inf_{\mu:\hat{\mu}_{0}=\nu,\hat{\mu}_{\infty}=\mu_{i,c}}\left(I_{{\rm eq},c}\left[\hat{\mu}_{0}\right]+J_{\left[0,\infty\right]}^{c}\left[\hat{\mu}\right]-I_{{\rm eq},c}\left[\hat{\mu}_{\infty}\right]\right)\\
&=&I_{{\rm eq},c}\left[\nu\right]-I_{{\rm eq},c}\left[\mu_{i,c}\right]+A_{i,c}\left[\nu\right],\label{rate function}
\end{eqnarray}
where $A_{i,c}\left[\nu\right]$ denotes a functional, 
\begin{equation}
A_{i,c}\left[\nu\right]=\left\{\begin{array}{ll}
0&{\rm if}\ \nu\ {\rm in}\ {\rm basin}\ {\rm of}\ \mu_{i,c}\\
{\rm positive}&{\rm others}
\end{array}
\right.
.\label{A}
\end{equation}
Notice that we do not need to reveal the explicit form of $A_{i,c}\left[\nu\right]$ to obtain the final result.

Let us give a nontrivial nonequilibrium work relation for metastable states by use of the above functionals. 
Suppose that the system is initially set in the $i$th metastable state, which differs from the initial condition of the ordinary Jarzynski equality. 
The expectation of the exponentiated work is then evaluated as, by using the same procedure as the above case for the Jarzynski equality, 
\begin{equation}
-\displaystyle \frac{1}{N}\log\left\langle e^{-N\beta w}\right\rangle_{i}=\inf_{\mu}\left(V_{i,\lambda_{0}}\left[\mu_{0}\right]+J_{\left[0,T\right]}\left[\lambda,\mu\right]+\beta w_{\left[0,T\right]}\left[\lambda,\mu\right]\right).
\end{equation}
By the symmetry (\ref{fluctuation theorem}) and equation (\ref{rate function}), we reach
\begin{eqnarray}
\displaystyle \nonumber-\frac{1}{N}\log\left\langle e^{-N\beta w}\right\rangle_{i}&=&\displaystyle \inf_{\mu}\{I_{\mathrm{e}\mathrm{q},\lambda_{0}}\left[\mu_{0}\right]+J_{\left[0,T\right]}\left[\lambda,\mu\right]+\beta w_{\left[0,T\right]}\left[\lambda,\mu\right]\\
\nonumber&&+A_{i,\lambda_{0}}\left[\mu_{0}\right]-I_{\mathrm{e}\mathrm{q},\lambda_{0}}\left[\mu_{i,\lambda_{0}}\right]\}\\
\nonumber&=&\left\{\beta\left(\phi_{{\rm eq}}\left(\beta,\lambda_{T}\right)-\phi_{{\rm eq}}\left(\beta,\lambda_{0}\right)\right)-I_{\mathrm{e}\mathrm{q},\lambda_{0}}\left[\mu_{i,\lambda_{0}}\right]\right\}\\
\displaystyle \nonumber&&+\inf_{\hat{\mu}}\left\{I_{\mathrm{e}\mathrm{q},\hat{\lambda}_{0}}\left[\hat{\mu}_{0}\right]+J_{\left[0,T\right]}\left[\hat{\lambda},\hat{\mu}\right]+A_{i\hat{,\lambda}_{T}}\left[\hat{\mu}_{T}\right]\right\}\\
\nonumber&=&\beta\left\{\phi_{{\rm eq}}\left(\beta,\lambda_{T}\right)-\phi_{i}\left(\beta,\lambda_{0}\right)\right\}\\
&&+\displaystyle \inf_{\hat{\mu}}\left\{I_{\mathrm{e}\mathrm{q},\hat{\lambda}_{0}}\left[\hat{\mu}_{0}\right]+J_{\left[0,T\right]}\left[\hat{\lambda},\hat{\mu}\right]+A_{i\hat{,\lambda}_{T}}\left[\hat{\mu}_{T}\right]\right\},\label{wr}
\end{eqnarray}
where we employ the fact that the free energy per degree of freedom of the $i$th metastable state \cite{12,07} can be represented as, by use of the rate function $I_{{\rm eq},\lambda}$ [see equation (B.11) in Appendix B], 
\begin{eqnarray}
\nonumber\phi_{i}\left(\beta,\lambda\right)&=&\epsilon_{\lambda}\left[\mu_{i}\right]-\left(1/\beta\right)s\left[\mu_{i}\right]\\
&=&\left(1/\beta\right)I_{{\rm eq},\lambda}\left[\mu_{i,\lambda}\right]+\phi_{{\rm eq}}\left(\beta,\lambda\right).
\end{eqnarray}
Notice that the infimum in equation (\ref{wr}), $\displaystyle \inf_{\hat{\mu}}\left(I_{{\rm eq},\hat{\lambda}_{0}}\left[\hat{\mu}_{0}\right]+J_{\left[0,T\right]}\left[\hat{\lambda},\hat{\mu}\right]+A_{i\hat{,\lambda}_{T}}\left[\hat{\mu}_{T}\right]\right)$, cannot be always equal to zero for any $\lambda$ in contrast with the case of the ordinary Jarzynski equality. 
However, due to the fact that each term in the infimum has a minimum which is zero separately, 
we can make the value of the infimum vanish by choosing external protocol $\lambda$ appropriately.
Taking into account the form of the functional $A_{i,c}\left[\nu\right]$ as in equation (\ref{A}), we find that the infimum can vanish without any additional information of the functional $A_{i,c}\left[\nu\right]$ if the time-reversal external protocol $\hat{\lambda}$ expresses the protocol which leads the system to the $i$th metastable state $\mu_{i,\hat{\lambda}_{T}}$ from the equilibrium state $\mu_{\mathrm{e}\mathrm{q},\hat{\lambda}_{0}}$  in thermodynamic limit. To be more precise, we explain this situation as follows. Consider a solution of the nonlinear diffusion equation (\ref{diffusion eq}) with the initial condition $\mu_{\mathrm{e}\mathrm{q},\hat{\lambda}_{0}}$ under the protocol $\hat{\lambda}$. Here, we denote this solution as $\mu_{t}^{*}\left[\hat{\lambda},\mu_{\mathrm{e}\mathrm{q},\hat{\lambda}_{0}}\right]$. If the solution at time $T,\ \mu_{T}^{*}\left[\hat{\lambda},\mu_{\mathrm{e}\mathrm{q},\hat{\lambda}_{0}}\right]$, is in the basin of the $i$th metastable state $\mu_{i,\hat{\lambda}_{T}}$, the infimum $\displaystyle \inf_{\hat{\mu}}\left(I_{{\rm eq},\hat{\lambda}_{0}}\left[\hat{\mu}_{0}\right]+J_{\left[0,T\right]}\left[\hat{\lambda},\hat{\mu}\right]+A_{i\hat{,\lambda}_{T}}\left[\hat{\mu}_{T}\right]\right)$ vanishes.
Accordingly, in the restricted case as mentioned above, we establish the nonequilibrium work relation for metastable states in the same form as the Jarzynski equality, which is equation (\ref{JE_meta}).

Before going to the conclusion, we provide a method to compute the free energy for the metastable state in the actual experiment by employing the nonequilibrium work relation obtained in the above discussion.
We first prepare an initial equilibrium state as depicted by I in Figure 2.
Suppose that we drive the system toward a specific metastable state ($i$ th state) we are interested in by controlling the external protocol $\lambda$ as II in Figure 2.
The measurement protocol starts from its metastable state.
We again drive the system by inversely controlling external protocol $\hat{\lambda}$, while measuring the performed work during II and III.
The final state denoted by III is not necessarily in equilibrium.
We then compute the free energy for the metastable state from the exponentiated work following equation (2).
\begin{figure}
\begin{center}
\includegraphics*[height=5cm]{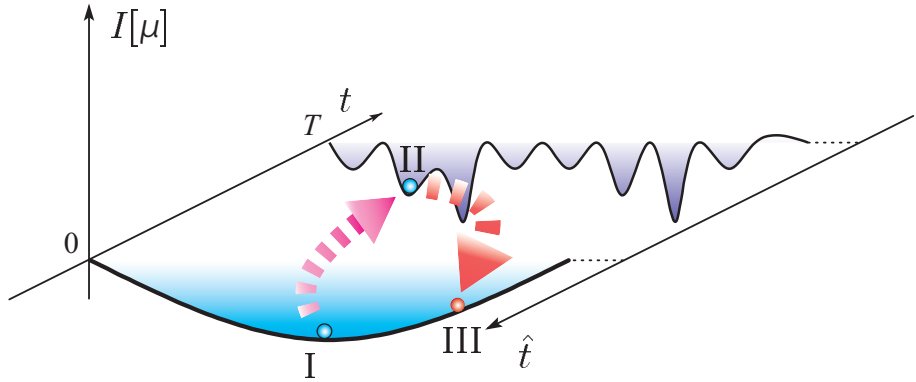}
\caption{(Color online) The lower parabolic curve denotes the trivial rate function for the initial condition.
The upper curve with many valleys represents the nontrivial rate function characterizing the interesting system with the metastable states.
The preparation of the experiment is as follows.
I. We prepare an initial equilibrium state.
II. We find a metastable state.
We record the protocol $\lambda$, which can lead to the metastable state we are interested in.
The main part of the experiment is from II to III.
III. We drive the system from the metastable state by the inverse protocol $\hat{\lambda}$, while we measure the performed work from II (red arrow).
}
\label{fig2}
\end{center}
\end{figure}

\section{Conclusion}
We have provided the correct understanding of the connection between the fluctuation theorem and the second law of thermodynamics, by employing the symmetry of the action functional.
This argument is succeeded in describing the behavior of the performed work in the thermodynamic limit. 
In addition, we have derived the nonequilibrium work relation for metastable states. 
As a result, this work relation can be expressed in the same form as the Jarzynski equality by the particular external protocol and enables us to compute the free energy for preselected metastable states by observing the performed work. 
Our result would serve as an important insight for the understanding of the peculiar relaxation observed as in glassy systems.

\ack
Y. S. acknowledges the financial support from JSPS Research Fellowships for Young Scientists and the Aihara Project, the FIRST program from JSPS, initiated by CSTP. 
M. O. thanks the hospitality of the Rome University during this work and financial support by MEXT in Japan, Grant-inAid for Young Scientists (B) No. 24740263.

\appendix
\section{}
In this appendix, we derive the functional Fokker-Planck equation (\ref{fokker-planck}) from the stochastic evolution of the density field, equation (\ref{dmu}). 
As an ordinary procedure to derive the Fokker-Planck equation from Ito process, we consider the derivative of test functional $\psi\left[\mu\right]$ as
\begin{eqnarray}
d\displaystyle \psi\left[\mu\right]=\int dx\frac{\delta\psi\left[\mu\right]}{\delta\mu\left(x\right)}d\mu\left(x\right)+\frac{1}{2}\int\int dxdy\frac{\delta^{2}\psi\left[\mu\right]}{\delta\mu\left(x\right)\delta\mu\left(y\right)}d\mu\left(x\right)d\mu\left(y\right).\label{A1}
\end{eqnarray}
By substituting equation (\ref{dmu}) into equation (\ref{A1}), we evaluate the expectation of test functional as
\begin{eqnarray}
\displaystyle \nonumber\left\langle d\psi\left[\mu\right]\right\rangle&=&\displaystyle \int dx\left\langle\frac{\delta\psi\left[\mu\right]}{\delta\mu\left(x\right)}D_{\mu,\lambda_{t}}\left(x\right)\right\rangle dt\\
\displaystyle \nonumber&&+\frac{1}{2N}\int\int dxdy\left\langle\frac{\delta^{2}\psi\left[\mu\right]}{\delta\mu\left(x\right)\delta\mu\left(y\right)}R_{\mu}\left(x,y\right)\right\rangle dt\\
\displaystyle \nonumber&=&\displaystyle \int dx\int\mathfrak{D}\mu\frac{\delta\psi\left[\mu\right]}{\delta\mu\left(x\right)}D_{\mu,\lambda_{t}}\left(x\right)\mathrm{Prob}_{t}\left[\mu\right]dt\\
\displaystyle \nonumber&&+\frac{1}{2N}\int\int dxdy\int\mathfrak{D}\mu\frac{\delta^{2}\psi\left[\mu\right]}{\delta\mu\left(x\right)\delta\mu\left(y\right)}R_{\mu}\left(x,y\right)\mathrm{Prob}_{t}\left[\mu\right]dt,\\\label{xxx66}
\end{eqnarray}
where $\mathrm{Prob}_{t}[\mu]$ denotes the probability for the density field. 
By integration by parts, equation (\ref{xxx66}) is calculated as 
\begin{eqnarray}
\displaystyle \nonumber\frac{\partial}{\partial t}\int\mathfrak{D}\mu\psi\left[\mu\right]\mathrm{Prob}_{t}\left[\mu\right]=\int\mathfrak{D}\mu\psi\left[\mu\right]\int dx\frac{\delta}{\delta\mu\left(x\right)}\left\{D_{\mu,\lambda_{t}}\left(x\right)\mathrm{Prob}_{t}\left[\mu\right]\right\}\\
\displaystyle \ \ \ \ +\int\mathfrak{D}\mu\psi\left[\mu\right]\frac{1}{2N}\int\int dxdy\frac{\delta^{2}}{\delta\mu\left(x\right)\delta\mu\left(y\right)}\left\{R_{\mu}\left(x,y\right)\mathrm{Prob}_{t}\left[\mu\right]\right\}.
\end{eqnarray}
Accordingly, we obtain the functional Fokker-Planck equation, 
\begin{eqnarray}
\displaystyle \nonumber\frac{\partial}{\partial t}\mathrm{Prob}_{t}\left[\mu\right]&=&\displaystyle \int dx\frac{\delta}{\delta\mu\left(x\right)}\left\{-D_{\mu,\lambda_{t}}\left(x\right)\mathrm{Prob}_{t}\left[\mu\right]\right\}\\
&&+\displaystyle \frac{1}{2N}\int\int dxdy\frac{\delta^{2}}{\delta\mu\left(x\right)\delta\mu\left(y\right)}\left\{R_{\mu}\left(x,y\right)\mathrm{Prob}_{t}\left[\mu\right]\right\}.
\end{eqnarray}
\section{}
In this appendix, we show that the rate function for the canonical distribution can be represented as $I_{{\rm eq},\lambda}\left[\mu\right]=\beta\epsilon_{\lambda}\left[\mu\right]-s\left[\mu\right]-\beta\phi_{{\rm eq}}\left(\beta,\lambda\right)$ in the case where the Hamiltonian of the system is described as equation (\ref{mfhm}). Here, $\epsilon_{\lambda},\ s$ and $\phi_{{\rm eq}}$ are denoted as follows.
\begin{eqnarray}
\displaystyle \epsilon_{\lambda}\left[\mu\right]=-\int F_{\lambda}\left(x\right)\mu\left(x\right)dx-\frac{1}{2}\int\mu\left(x\right)G\left(x,y\right)\mu\left(y\right)dxdy,\label{b1}\\
s\displaystyle \left[\mu\right]=-\int\mu\left(x\right)\log\mu\left(x\right)dx,\label{b3}\\
\displaystyle \phi_{{\rm eq}}\left(\beta,\lambda\right)=-\frac{1}{N\beta}\log Z_{\beta,\lambda,N},\label{b4}
\end{eqnarray}
where $Z_{\beta,\lambda,N}$ denotes the partition function. 
Let us consider the coarse grained picture of the canonical distribution by the density field $\mu$.
The canonical distribution is represented as  
\begin{equation}
\displaystyle \rho_{\beta,\lambda,N}\left(\left\{x_{i}\right\}\right)\prod_{i}dx_{i}=\frac{1}{Z_{\beta,\lambda,N}}e^{-\beta H_{N}\left(\left\{x_{i}\right\},\lambda\right)}\prod_{i}dx_{i}.\label{canonical distribution}
\end{equation}
We then obtain the coarse grained distribution of $\mu$ as
\begin{eqnarray}
\displaystyle \nonumber \mathrm{Prob}_{\beta,\lambda}\left[\mu\right]\mathfrak{D}\mu&=&\displaystyle \frac{1}{Z_{\beta,\lambda,N}}e^{-N\left\{-\beta\int F_{\lambda}\left(x\right)\mu\left(x\right)dx-\frac{\beta}{2}\int\mu\left(x\right)G\left(x,y\right)\mu\left(y\right)dxdy\right\}}W\left[\mu\right]\mathfrak{D}\mu.\\\label{B2}
\end{eqnarray}
Here, $W\left[\mu\right]$ denotes the state density satisfying $\Pi_{i}dx_{i}=W\left[\mu\right]\mathfrak{D}\mu$ and we employed the fact that the Hamiltonian was described by use of $\mu$ as
\begin{equation}
H_{N}\displaystyle \left[\mu\right]=-N\int F_{\lambda}\left(x\right)\mu\left(x\right)dx-\frac{N}{2}\int\mu\left(x\right)G\left(x,y\right)\mu\left(y\right)dxdy.
\end{equation}
According to the Sanov's theorem [10], the rate function of the density field for IID random variables with a common probability density $\rho\left(x\right)$ is given as $I\left[\mu\right]= \int dx\mu\left(x\right)\log\left(\mu\left(x\right)/\rho\left(x\right)\right)$. 
Here, suppose that the common probability density is the uniform density $\rho\left(x\right)=1/\Lambda$, where $\Lambda$ denotes the cardinality of domain for $x$; we then obtain the probability density of the density field as
\begin{equation}
\mathrm{Prob}[\mu]=\exp\left[-N\int dx\mu\left(x\right)\log\mu\left(x\right)-N\log\Lambda\right].
\end{equation}
On the other hand, by using the state density, we can express $\mathrm{Prob}[\mu]$ as 
\begin{equation}
\displaystyle \mathrm{Prob}[\mu]=\frac{W\left[\mu\right]}{\Lambda^{N}}.
\end{equation}
By comparing equation (B.7) to equation (B.8), we can evaluate the state density in a sufficient large $N$ as
\begin{equation}
W\left[\mu\right]=e^{-N\int\mu\left(x\right)\log\mu\left(x\right)dx}.\label{state density}
\end{equation}
By substituting equation (\ref{state density}) into equation (\ref{B2}) and taking equations (\ref{b1}), (\ref{b3}) and (\ref{b4}) into account, we obtain 
\begin{equation}
\mathrm{Prob}_{\beta,\lambda}\left[\mu\right]=e^{-N\left\{\beta\epsilon_{\lambda}\left[\mu\right]-s\left[\mu\right]-\beta\phi_{{\rm eq}}\left(\beta,\lambda\right)\right\}}.
\end{equation}
Accordingly, we find the rate function of the canonical distribution as
\begin{equation}
I_{{\rm eq},\lambda}\left[\mu\right]=\beta\epsilon_{\lambda}\left[\mu\right]-s\left[\mu\right]-\beta\phi_{{\rm eq}}\left(\beta,\lambda\right).
\end{equation}



\section*{References}
\begin{thebibliography}{00}

\bibitem{01}            Jarzynski C, 1997 Phys. Rev. Lett. {\bf 78} 2690; 1997 Phys. Rev. E {\bf 56} 5018

\bibitem{02}            Jarzynski C, 2000 J. Stat. Phys. {\bf 98} 77

\bibitem{03}            Crooks G E, 1999 Phys. Rev. E {\bf 60} 2721

\bibitem{14}            Crooks G E, 2000 Phys. Rev. E {\bf 61} 2361

\bibitem{15}            Seifert U, 2005 Phys. Rev. Lett. {\bf 95} 040602

\bibitem{16}            Jarzynski C, 1999 J. Stat. Phys. {\bf 96} 415

\bibitem{04}            Chetrite R and Gaw\c{e}dzki K, 2008 Commum. Math. Phys. {\bf 282} 469

\bibitem{18}            Harris R J and Sch\"{u}tz G M, 2007 J. Stat. Mech. P07020

\bibitem{17}            Ellis R S, 1985 {\it Entropy, Large Deviations, and Statistical mechanics} (Springer)

\bibitem{12}            Touchette H, 2009 Phys. Rep. {\bf 478} 1

\bibitem{06}            Dawson D A, 1983 J. Stat. Phys. {\bf 31} 29

\bibitem{07}            Dawson D A and G\"{a}rtner J, 1989 Mem. Am. Math. Soc. {\bf 78} 398

\bibitem{10}            Freidlin M I and Wentzell A D, 1984 {\it Random Perturbations of Dynamical Systems} (Springer)

\bibitem{Ohzeki2012}   Ohzeki M, 2012 Phys. Rev. E  {\bf 86} 061110

\bibitem{05}            Gardiner C W, 1985 {\it Handbook of Stochastic Methods} (Springer, 2nd Edition)

\bibitem{08}            Shiino M, 1987 Phys. Rev. A {\bf 36} 2393

\bibitem{09}            Dean D S, 1996 J. Phys. A: Math. Gen. {\bf 29} L613

\bibitem{11}            Bertini L, Sole A D, Gabrielli D, Jona-Lasinio G and Landim C, 2001 Phys. Rev. Lett. {\bf 87} 040601; 2002 J. Stat. Phys. {\bf 107} 635

\bibitem{13}            Varadhan S R S, 1966 Comm. Pure. Appl. Math. {\bf 19} 261

\end{thebibliography}
\end{document}